\documentstyle[12pt]{article}
\begin{document}

\title{\bf Green's Function Approach to B. E. Condensation in the alkali atoms }
\author{Mahendra Sinha Roy }
\maketitle
\centerline{\small Department of Physics, Presidency College, Calcutta 700 073,India.}
\date{}
\vspace{.5 cm}

\begin{abstract}
{\it A new  scheme has been proposed to solve the B. E. condenstates  in terms of Green's function  approach. It has been shown that the radial wave function of two interacting atoms, moving in a common harmonic oscillator potential modified by an effctive interaction, satisfies an intregral equation whose kernel is separable. The solution of the integral equation can be written in terms of the harmonic oscillator wave functions.  The ground state wave function of the system can be written in terms of these solutions.}

\end{abstract}
\newpage

The recent experimental realization of Bose-Einstein condensation [BEC] in ultra cold atomic gases has generated intensive theoretical and practical interest regarding the properties of Bose gas. Non-linear Schrodinger equation in this context known as Gross-Pitaevskii [ G-P ] equation [1],has been used to study the properties of the interacting spinless Bosons , trapped in harmonic potential . Altough the G-P equation  is widely accepeted as a workable model for description of the B E condenstates at low temperature, the actual dynamics of B E condenstates is not fully captured by the G-P equation because of the non-integrability of the G-P equation and explicit solution is not known [2]. It is to be noted that for the positive scattering length, the G-P equation is generally solved numerically only under cylindrically symmetric systems. The Thomas Fermi approximation [3] has been introduced to get some analytical solution for the description of B E condenstates. On the other hand for the negative scattering length the G-P equation is yet to be properly investigated [3]properly. Another alternative approach to study the dynamics of the B.E.condenstates is the time dependent variation technique   [4] in which one assumes fixed profile of the condenstate and computes the evolution of some parameters such as the width by variational techniques. It should also be mentioned that the negative scattering length provides stable solution of the G-P equation only under certain conditions for the number of particles and the size of the trap [5]. When those  conditions are not fulfiled the condenstates become unstable. Therefore realization of large condenstates will provide reliable experimental data for the precise understanding of the condensation process[6].
In this paper we shall present a new solution for the condenstate on the basis of Green's function  approach of scattering theory. Similar method was used in the treatment of the nuclear many body problems by Bethe and Goldstone; Brueckner and Gammel [7] for the understanding of the various aspects nuclear many body problem. We have derived an integral equation for the description of two interacting bosons in the presence of many bosons. 
We shall  consider a system of N spinless interacting atoms all having the  same hyperfine species, in equilibrium and at very low temperature and densities characteristic of alkali atoms under Bose- Einstein condensation (BEC) condition. We assume that each of the atoms is moving in the influence of a common harmonic oscillator potential which is called a trapping potential($V_{ext}$) and an effective two-body interaction acting between them which modifies their behaviour . The essential questions we wish to investigate are : How does the effective two body interaction modify the two-body wave function?. How does the energies of the two interacting bosons system change when the  two- body effective interaction is switched on ?  It is welknown that the scattering length ($a_s$) is very small compared to the interatomic distant($r_{int}$) [8] under BEC in the dilute atomic alkali gases. Therefore the behaviour of the atoms is extremely sensitive to the details of the trap and the effective two- body interaction. Since the effective two- body interaction is very weak the quantitative calculation based on perturbation theory in the interatomic interaction is hoped to be highly reliable for the alkali gases. Furthermore the de Broglie wave length associated with individual atom is comparable to interatomic $r_{int}$ distance therefore the correlation between atoms becomes important[7]. Hence the construction of the  many -body condenstate state wave function of the system in terms of two-body wave function will be interesting. The standard many-body wave function in terms of Hartree-Fock ansatz  using the single  particle solution of the G-P equation does not take into accoutn the effect of correlation between atoms which is very important in the case of BEC alkali gases.In this letter we have presented an analytical formalism to investigate the above aspects of BEC problem  without any numerical estimate for a particular choice of the effective interaction.That will be   communicated shortly.

 Let us consider a system of interacting $N$ spinless atoms and now arbitrarily pick out any two interacting spinless atoms, say 1 and 2, denoting their relative and center-of- mass co-ordinates, by $\vec r $ and $\vec R $  respectively . We shall investigate the state ($\phi$) of these two interacting atoms trapped in a common harmonic potential in the presence $(N-2)$ interacting atoms. The most general states of the many-body system will then be described by a wave function $\Psi(\vec r,\vec R,{\chi},t)$ where ${\chi}$ represents schemitically the co-ordinates of the remaining $(N-2)$ particles. It is the dependence on $\vec r$ that is crucial but the parameter dependence of $\Psi$ on $\vec R , \chi$ and $t$ will be ignored . In the present context we assume that simple BEC is realised in the two-particle state ${\phi(\vec r)}$ whose equatiom of motion will be determined using Schrodinger equation. From the knowledge of the wave-function of two-atom system ${\phi(\vec r)}$, the many- body ground state of B E condenstate system of  alkali atoms can be written. The hamiltonian for the two- atom system in the common harmonic oscillator potential of frequency $\omega$ is given by 

\begin{equation}H_0=\frac{p_1^2}{2m}+\frac{p_2^2}{2m}+ \frac{m\omega^2r_1^2}{2}+ \frac{m\omega^2r_2^2}{2}.\end{equation}
Introducing the cannonical co-ordinate transform
\begin{equation}\vec r=\vec r_2-\vec r_1  ,  \vec p=\frac{(\vec p_2-\vec p_1)}{2},\end{equation} and 

\begin{equation}\vec R=\frac{(\vec r_2+\vec r_1)}{2},\vec P=(\vec p_2+\vec p_1).\end{equation}
The hamiltonian is transformed to
\begin{equation}H_0=\frac{(P^2+M^2\omega^2R^2)}{2M} +\frac{(p^2+\mu^2 \omega^2 r^2)}{2\mu},\end{equation}
where M=2m,$\mu=\frac{m}{2}$.
The Schrodinger equation for the system  with $\hbar=1 $ is
\begin{equation}[\frac{(-\nabla_R^2+M^2\omega^2R^2)}{2M}+\frac{(-\nabla_r^2+\mu^2\omega^2r^2)}{2\mu}]\phi(\vec r)\phi(\vec R)=E^0 \phi(\vec r)\phi(\vec R).\end{equation}
The Schrodinger equation for the relative motion is then
\begin{equation}H_0\phi(\vec r)=(-\frac{\nabla^2}{2\mu}+k r^2)\phi(r)=E^0\phi(\vec r),\end{equation}
where \begin{equation}k=\frac{\mu \omega^2}{2}=\frac{m \omega^2}{4}.\end{equation}

The solutions are given by
\begin{equation}\phi_{nl}^m(\vec r)=\frac{R_{nl}(r)}{r}Y_{lm}({\hat{ r}}),\end{equation} where $Y_{lm}({\hat {r}})$ are the usual spherical harmonics.
The radial wave function $R_{nl}(r)$ is given by 
\begin{equation}R_{nl}(r)=N_{nl}exp(-\frac{\nu r^2}{2})r^{l+1}v_{nl}(r),\end{equation}
where $\nu = \frac{\omega\mu}{\hbar}$ and $v_{nl}(r)$ is the associated Laguerre polynomial:
\begin{equation}v_{nl}(r)=L_{n+l+\frac{1}{2}}^{l+\frac{1}{2}}(\nu r^2).\end{equation}
 $N_{nl}$ is determind by  the normalization  condition and $E_{nl}^{0}=\hbar\omega(2n+l+\frac{3}{2})$.

We shall now treat the effective spin independent two-body interaction $v(r)$ which is very weak between two atoms , as a peturbation and solve the Schrodinger equation using the Green's function method to determine the perturbed wave function for the two- atom system.
The unperturbed wave function is given by
\begin{equation}\phi_{nl}(\vec r)=\frac{R_{nl}(r)}{r}Y_{lm}(\hat{ r}).\end{equation}
Similarly the perturbed wave function can be written as
\begin{equation}\psi_{n_1l_1}(r)=\frac{u_{n_1l_1}}{r}Y_{l_1m_1}(\hat{ r}).\end{equation}
The Schrodinger equation for the perturbed system is
\begin{equation}[H_0+v(r)]\frac{u_{n_1l_1}}{r}Y_{m_1l_1}(\hat{ r})=E_{n_1l_1}\frac{u_{n_1l_1}}{r}Y_{m_1l_1}(\hat{r}),\end{equation}
where $E_{n_1l_1}$ is the perturbed energy eigen value.
We now write
\begin{equation}H_0\frac{R_{nl}(r)}{r}Y_{lm}(\hat{ r})=\frac{1}{r}[-\frac{1}{m}\frac{\partial^2}{\partial r^2}+V_l(r)]R_{nl}(r)Y_{nl}(\hat{ r})=E_{nl}^0\frac{R_{nl}(r)}{r}Y_{nl}(\hat{r}).\end{equation}
Multiplying the perturbed equation(13) by $Y_{n_1l_1}^*(\hat{r})$ on the left and integrating over the solid angle $d\hat{r}$ we obtain
\begin{equation}[E_{n_1l_1}-(-\frac {1} {m}\frac{d^2}{dr^2}+V_{l_1}(r)]u_{n_1l_1}(r)=v_{l_1}(r)u_{n_1l_1}(r),\end{equation}
where\begin{equation}v_{l_1}(r)=\int d\vec rY_{l_1m_1}^*(\hat{r})v(r)Y_{l_1m_1}(\hat{r})\end{equation}
In order to solve this equation we introduce Green`s $G_{n_1l_1'}(\vec r,\vec r')$ corresponding to equ.(15) satisfies the equation
\begin{equation}(E_{n_1l_1}-H_0)G_{n_1l_1}(\vec r,\vec r')=\delta(\vec r-\vec r`).\end{equation}
Multiplying through on the left by $Y_{l_1m_1}^*(\hat{r})$ and integrate over the solid angle $dr$ we obtain
\begin{equation}[E_{n_ll_1}-(-\frac{1}{m}\frac{d^2}{dr^2})+V_{l_1}(r)]g_{n_1l_1}(r,r')=\delta(r-r'),\end{equation}
where \begin{equation}g_{n_1l_1}(r,r')=\sum_{n_2}{\frac{R_{n_2l_1}(r)R_{n_2l_1}(r')}{E_{n_1l_1,n_2l_1}}},\end{equation}
where
\begin{equation}E_{n_1l_1,n_2l_2}=(E_{n_1l_1}-E_{n_2l_1}^0).\end{equation}

Hence the perturbed wave function $u_{n_1l_1}(r)$ satisfies an integral equation with a separable kernel. Therefore we write
\begin{equation}u_{n_1l_1}(r)={\int_0^{\infty} dr' g_{n_1l_1}(r,r')v_{l_1}(r')u_{n_1l_1}(r')}.\end{equation}
Finally we write
\begin{equation}u_{n_1l_1}(r)=\sum_{n_2 =0}^{\infty}{ \frac{R_{n_2l_1}(r)}{E_{n_1l_1},n_2l_1}\int_0^{\infty} dr' R_{n_2l_1}(r')v_{l_1}(r')u_{n_1l_1}(r')}.\end{equation}
The equation for $R_{nl_1}(r)$ is
\begin{equation}[E_{nl_1}^0-(-\frac{1}{m}\frac{d^2}{dr^2}+V_l{_1}(r))]R_{nl_1}(r)=o.\end{equation}
Multiplying equ.(15)and equ.(23) on the left by $R_{nl_1}(r)$ and $u_{n_1l_1}$respectively subtracting and integrating over $r'$ we obtain
\begin{equation}E_{n_1l_1}-E^0_{nl_1}=\frac{\int_0^{\infty} dr {R_{nl_1}(r)v_{l}(r)u_{n_1l_1}(r)}}{\int_0^{\infty} dr R_{nl_1}(r)u_{n_1l_1}(r)}\end{equation}

and the perturbed wavefunction $u_{n_1l_1}(r)$ in a closed form can be written as
\begin{equation}u_{n_1l_1}(r)=\sum_{n_2=0}^{\infty}{K_{n_1l_1,n_2l_1}R_{n_2l_1}( r)},\end{equation}

where \begin{equation}K_{n_1l_1,n_2l_1}={\frac{1}{E_{n_1l_1,n_2,l_1}}\int_0^{\infty}dr R_{n_2l_1}(r)v_{l_1}(r)u_{n_1l_1}(r)}.\end{equation}.  To this expression of $u_{n_1l_1}$ we should add the solution of $(E_{n_1l_1}-H_0){\chi}_{n_1l_1}=0$. Since $E_{n_1l_1}\ne E^0_{nl_1}$ the equation $(E_{n_1,l_1}-H_0){\chi}_{n_1l_1}=0$ has no solution.Thus it is not necessary to add a particular solution to the $u_{n_1l_1}$ . From equations(25) and (26)it follows that
\begin{equation}K_{n_1l_1,n_2l_1}=\frac{1}{E_{n_1l_1,n_2l_1}}\sum_{n_3=0}^{\infty}{K_{n_1l_1,n_3l_1}\int_0^{\infty} dr R_{n_2l_1}(r)v_{l_1}(r)R_{n_3l_1}(r)}.\end{equation}.

\begin{center}
{\bf Results and Discussion}
\end{center}
The main conclusions of the paper are the following:
It should be noted that $n_3$ can in principle run from $o $ to $ \infty $, the above equ.(27) reprsents an infinite set of simultaneous homogeneous algebraic equations in $K_{n_1l_1,n_2l_1}$. The consistency of the these equations imposes the condition of vanishing of the  determinant formed from the coefficients of $K_{n_1l_1,n_2l_1}$. This condition determines  the energy shift $E_{n_1l_1,nl_1}$of the two interacting atoms. The corresponding eigenvectors when substituted in eq.(26) determine the perturbed radial function $u_{n_1l_1}(r)$. Since B E condensation takes place strictly in  the ground state of the system therefore a few values of $n_3$ in the sum will provide  a very reliable  estimate of the two-particle wave function. Because of the separable structure of eq.(21) we can also include various types of central and non central  of two-body effective interaction  and the variation of the scattering length can also be incorporated to investigate sensitivity of the two-body condenstate wave function and the stability of the Bose-Einstein condensation in alkali atoms on the effective two-body interaction which can not be so easily  taken care in the formalism of G-P equation. When the effective two-body interaction becomes non-central in nature then the radial equation for  $u_{n_1l_1}(r)$ i.e eq.(21) will satisfy a coupled integral equation. When the effective two-body interaction is a pseudo- potential type, as commonly used in the calculation of G-P equation, the determination of the two-body wave function $u_{n_1l_1}(r)$ becomes extermely simple in this formalism. We can also investigate the importance two- particle correlation in BEC. Finally the construction of the  condenstate wave function describing 
the BEC of the in terms of the calculated two- particle wave function to study various aspects of BEC becomes possible. These are the  main conclusions of the paper.

\end{document}